\documentclass[showpacs,twocolumn,amssymb,aps,flushbottom,floatfix,balancelastpage,superscriptaddress]{revtex4-2}
\usepackage{graphicx}
\usepackage{amsmath}
\usepackage[colorlinks=true]{hyperref}
\usepackage{subfig}
\usepackage{natbib}
\usepackage{enumitem}
\usepackage{ulem}

\begin{document}


\title{Higher-order rogue wave fission under the effects of third-order dispersion, self-steepening and self-frequency shift}

\author{Amdad Chowdury}
\affiliation{School of Physical and Mathematical Sciences, Nanyang Technological University, 637371, Singapore}

\author{Wonkeun Chang}
\affiliation{School of Electrical and Electronic Engineering, Nanyang Technological University, 639798, Singapore\looseness=-1}

\author{Marco Battiato}
\affiliation{School of Physical and Mathematical Sciences, Nanyang Technological University, 637371, Singapore}

\begin{abstract}
Using the generalised nonlinear Schr\"odinger equation, we investigate how the effect of third-order dispersion, self-steepening, and Raman-induced-self-frequency shift have an impact on the higher-order rogue waves. We observe that individually each effect breaks apart the higher-order rogue waves reducing to their constituent fundamental parts similar to how a higher-order soliton undergoes a fission. We demonstrate that under the influence of their combined effect, the disintegrated elements of higher-order rogue waves becomes fundamental solitons creating an asymmetrical spectral profiles that generates both red and blue-shifted frequency components. These observations reveal the mechanisms that create a large number of solitons in the process of modulation instability-induced supercontinuum generation from a continuous-wave background in optical fibers.
\end{abstract}

\maketitle

\section{Introduction}
\label{one}
The rogue wave solution of the nonlinear Schr\"odinger equation (NLSE) is attractive because of its ability to explain sudden extreme formation in nature such as oceanic rogue waves, shaping of light pulse with unusually high intensity and creation of localised structures in Bose-Einstein condensates etc. 
After three decades of intense study, the rogue wave solution has extended to a range of other multidisciplinary fields from optics \cite{Akhmediev:09:PLA1,onorato2001freak,muller2005rogue,Kibler:10:Nature,moslem2011surface,stenflo2010rogue,Efimov:10:EPJST,Bludov:10:EPJST,Shats:10:PRL,veldes2013electromagnetic,moslem2011dust,tsai2016generation,bludov2009matter} to economics \cite{zhen2010financial} . 

In nonlinear optics, the formation of optical rogue waves is a short burst of light pulses with high intensities that appears in a chaotic optical wave-field. The first demonstration of the existence of an optical rogue wave was reported in the work of Solli et.al. \cite{Solli:07:Nature} in fiber supercontinuum generation. Since then, the phenomenon has been studied in  various different branches of nonlinear optics because of their multiplicity of applications and impact such as in optical cavities \cite{montina2009non,residori2012rogue}, mode-locked lasers \cite{lecaplain2012dissipative,soto2011dissipative}, photonic crystal fibers \cite{buccoliero2011midinfrared}, Raman fibers laser and amplifiers \cite{runge2014raman,finot2009selection} and optical parametric process \cite{hammani2009emergence}. A comprehensive report on recent progress in research of optical rogue waves in the field of nonlinear optics can be found in these works \cite{akhmediev2013recent,song2020recent}.

Instability seeded by noise acts as a breeding ground for the emergence of optical rogue waves. This is described by a process called modulation instability (MI). It is a complex nonlinear process which can exponentially amplify a small noise or disturbance leading to a drastic change in the system \cite{zakharov2009modulation} such as forming optical rogue waves in optical fibre. 

Despite the strong interest on the nature of optical rogue wave and how it emerges and influences an optical system, many of its characteristics are not well understood. For instance, during propagation of ultrashort optical pulse through an optical fiber, the fibre medium exerts higher-order linear and nonlinear effects. Under these effects, how the spontaneous MI develops, where it leads to and the impact of MI on a particular optical phenomena is not well understood. One prime example where we can observe the MI which plays the pivotal role is in the process of continuous wave supercontinuum generation (CW-SCG). 

In the CW-SCG regime, the continuous wave pump can be considered as a higher-order soliton of a very large soliton number. Their evolution is dominated by a noise-seeded MI leading to its break-up \cite{super}. The disintegration is highly non-trivial and is a multi-stage process. At the beginning of the evolution, the presence of noise among the large number of solitons produces MI, making the higher-order soliton unstable. Immediately before the higher-order soliton collapses, this MI is strong enough to form a wide variety of rogue waves-type substructures. In the final stage, all these rogue waves become a collections of fundamental solitons. However, the physics behind how the initial MI results in the production of many solitons is relatively less studied, and a definitive insight into the process is yet to be realized.

Applying the NLSE, the formation of fundamental and higher-order rogue waves as a result of noise driven MI on a continuous wave field is already presented in \cite{toenger2015emergent}. In this work, using the generalised nonlinear Schr\"odinger equation (GNLSE), we show that higher-order rogue waves undergo fission, similarly to higher-order soliton fission. The GNLSE is perturbed with third-order dispersion (TOD), self-steepening (SS), and Raman induced self-frequency shift (RIFS) effects. We solve it numerically including the TOD, SS, and RIFS individually and show that each of these effects induce rogue wave fission breaking the higher-order rogue wave to their constituent parts. Note that while all these effects induce fission in the higher-order rogue waves, as we shall see later, it is the RIFS effect that has the major impact on transforming the disintegrated rogue waves to solitons. Finally we solve the full GNLSE and demonstrate that simultaneously, these three effect induce fission on the higher-order rogue waves and after a long-time evolution the breakaway rogue wave components transform to fundamental solitons. 

The article is organized as follows. In Sec.(\ref{three}), (\ref{four}), and (\ref{five}), we investigate the effect of TOD, SS, and RIFS on a second and third-order rogue waves showing that each of them individually induce rogue wave fission. We also demonstrate that under the RIFS effect, higher-order rogue waves generate red-shifted frequency components while decelerating, until eventually transforming into a soliton. In Sec.(\ref{six}), we investigate the combined influence, showing that after the occurrence of the fission, the evolving rogue wave components create an asymmetrical spectral profiles producing both red and blue shifted frequencies.

\subsection {Model, solutions and techniques}
The GNLSE in its normalized form is
\begin{multline}
\label{gnlse}
i \frac{\partial \psi}{\partial z}-\frac{\beta_2}{2} \frac{\partial^2 \psi}{\partial t^2}+\gamma \,\psi\lvert\psi\rvert^2=\\
 i\epsilon_3\, \frac{\partial^3 \psi}{\partial t^3}-is\frac{\partial}{\partial t}(\psi\lvert\psi\rvert^2)+\tau_R \psi \frac{\partial|\psi|^2}{\partial t}\textrm{,}
\end{multline}
where $\psi=\psi(z,t)$ is the complex field envelop, $z$ the evolution variable and $t$ the transverse variable. $\epsilon_3$ is the TOD parameter,  $s$ and $\tau_R$ are the coefficients of SS and RIFS with their explicit expression given by
\begin{equation}\label{parameterss}
\epsilon_3 =   \frac{\beta_3}{6|\beta_2|\,t_0}, \; s=\frac{1}{\omega_0\,t_0}, \; \tau_R=\frac{T_r}{t_0}
\end{equation}
where $\beta_2$ is the group velocity dispersion (GVD), $\gamma$ the nonlinear strength, $\beta_3$  the coefficient of TOD, $\omega_0$ the carrier angular frequency, $t_0$  the pulse duration, and $T_r$  the Raman time constant \cite{atieh1999measuring}. Since the coefficients of TOD, SS and RIFS in Eq.(\ref{gnlse}) are inversely proportional to the pulse duration $t_0$, they are negligible in the long pulse regime, while they contribute significantly in the ultrashort pulse regime. 

With GVD parameter $\beta_2=-1$, nonlinear parameter $\gamma=1$ and $\epsilon_3=s=\tau_R=0$, the Eq.(\ref{gnlse}) is the NLSE and is the most basic form of the equation that can be used to model optical pulse propagation in nonlinear dispersive media \cite{agrawal2011nonlinear}. This form of the equation can be solved analytically using the inverse scattering transformation \cite{Zakharov:72:JETP}. In this work, we assume that the TOD, SS and RIFS effects are perturbations to the rogue wave solution of NLSE. We apply small magnitude of perturbation and solve Eq.(\ref{gnlse}) numerically.

To numerically generate higher-order rogue waves, we adopt the analytic solutions as the initial conditions well before their fully developed stage. Specifically we used the second and third-order rogue wave solution as presented in \cite{Akhmediev:09:PRE}(see eq.~23 and 26). The analytic solutions in their exact form can be obtained by solving NLSE through the Darboux transformation technique using a continuous wave, $\psi=\exp\left(iz\right)$, as the seed \cite{Matveev:91:Book}. The main features of a higher-order rogue wave of order $N$ is, they consists of $N (N + 1)/2$ fundamental rogue waves and can reach the maximum amplitude $2N +1$. The phase shift across the peak becomes $\pi$. The details of the employed numerical techniques can be found in \cite{chowdury2021rogue}.

Apart from the investigation of higher-order rogue waves fission, we also study their impact on the surrounding wave background. In optics or hydrodynamics, in a turbulent wave field a variety of waves can be formed. Recently, the inverse scattering techniques (IST) has been successfully applied to classify them. We will leverage this technique to identify particular types of wave from  calculated spectral profiles. The details of the technique and its implementation is outlined in \cite{randoux2016inverse,randoux2018nonlinear,bonnefoy2020modulational}. 

All simulations throughout this work range from $z=0$ till $z=24$ will be ignored. However figures will show only the fraction of the calculated range which shows rogue waves' structures and interactions.

\section{Effect of TOD} 
\label{three}

\begin{figure}[htbp]
\centering
\includegraphics{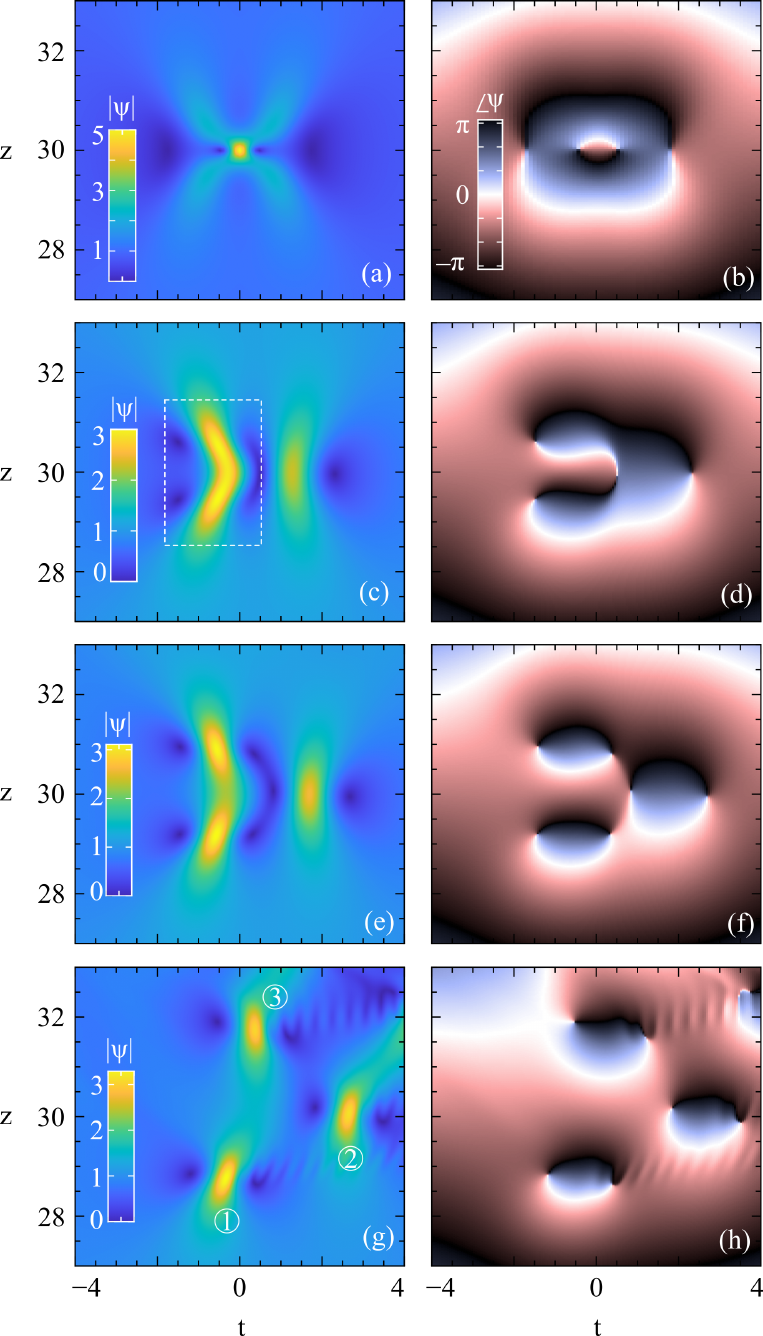}
\caption{ (a) Amplitude and (b) phase profiles of the second-order rogue wave with $\beta_3=0$. Panels (c) and (d) ($\beta_3=0.04$) show how the rogue wave breaks apart into a doublet and a first-order rogue wave. Panels (e) and (f) ($\beta_3=0.08$) show  the doublet further breaking into two conjoined first-order rogue waves. Panels (g) and (h) are with $\beta_3=0.2$ where each of the rogue waves are well separated.}
\label{todsecond}
\end{figure}

We first address only the TOD term with $\beta_2=-1, \gamma=1$ and $s=\tau_R=0$ in Eq.~\ref{gnlse}. The modified equation is  given by
\begin{equation}
\label{nlse-tod}
i \frac{\partial \psi}{\partial z}+\frac{1}{2} \frac{\partial^2 \psi}{\partial t^3}+\psi\lvert\psi\rvert^2-i\epsilon_3\, \frac{\partial^3 \psi}{\partial t^3}=0\textrm{.}
\end{equation}
Since no analytic solution is known, the effect of the perturbation on the second-order rogue wave is examined numerically by solving Eq.~(\ref{nlse-tod}) for a range of $\beta_3$ values (and therefore TOD parameter $\epsilon_3$, as in Eq.~\ref{parameterss}). 
The temporal and phase profiles of the numerically generated second-order rogue wave is shown in Fig.~\ref{todsecond}(a) and \ref{todsecond}(b). With $\beta_3=0$, the maximum height reached to $5$ while there is a $\pi$ phase shift at $z=30$ across the transverse direction. 
 
An in-depth analysis of the fission under TOD is described in \cite{chowdury2021rogue}. Here we focus on how increasing magnitude of TOD coefficient $\beta_3$ affects the higher-order rogue waves. We simulated a second-order rogue wave with a range of $\beta_3$ values starting from $0.01$ to $0.2$. Results are presented in in Fig.~\ref{todsecond}, showing the onset of the fission process.

When the TOD term is small ($\beta_3=0.04$), we observe already a collapse of the rogue wave. The main structure separates into two substructures, creating a doublet accompanied by one separate first-order rogue wave. The doublet within the white rectangle in Fig.~\ref{todsecond}(c) appeares as two conjoined first-order rogue waves with amplitude $3$ while a separate first-order rogue wave appears on the right side of the doublet. Corresponding phase profiles are shown in Fig.~\ref{todsecond}(d).

Interestingly, just a slightly higher values of $\beta_3=0.08$ leads to the breaking of the doublet, that becomes two independent first-order rogue waves. Including the first-order rogue wave in the right, there are three rogue waves appearing in Fig.~\ref{todsecond}(e). Independently each rogue wave's amplitude is $\approx 3$ and they undergo a $\pi$ phase shift shown in Fig.~\ref{todsecond}(f). As  $\beta_3$ increases, the three rogue waves get pushed further apart. For instance, with $\beta_3=0.2$ the three first-order rogue waves (marked by numbers and presented in Fig.~\ref{todsecond}(g)) are completely separated and appear independently at different space and time, while undergoing a $\pi$ phase shift, as observed in Fig.~\ref{todsecond}(h).

\begin{figure}[h]
\centering
\includegraphics{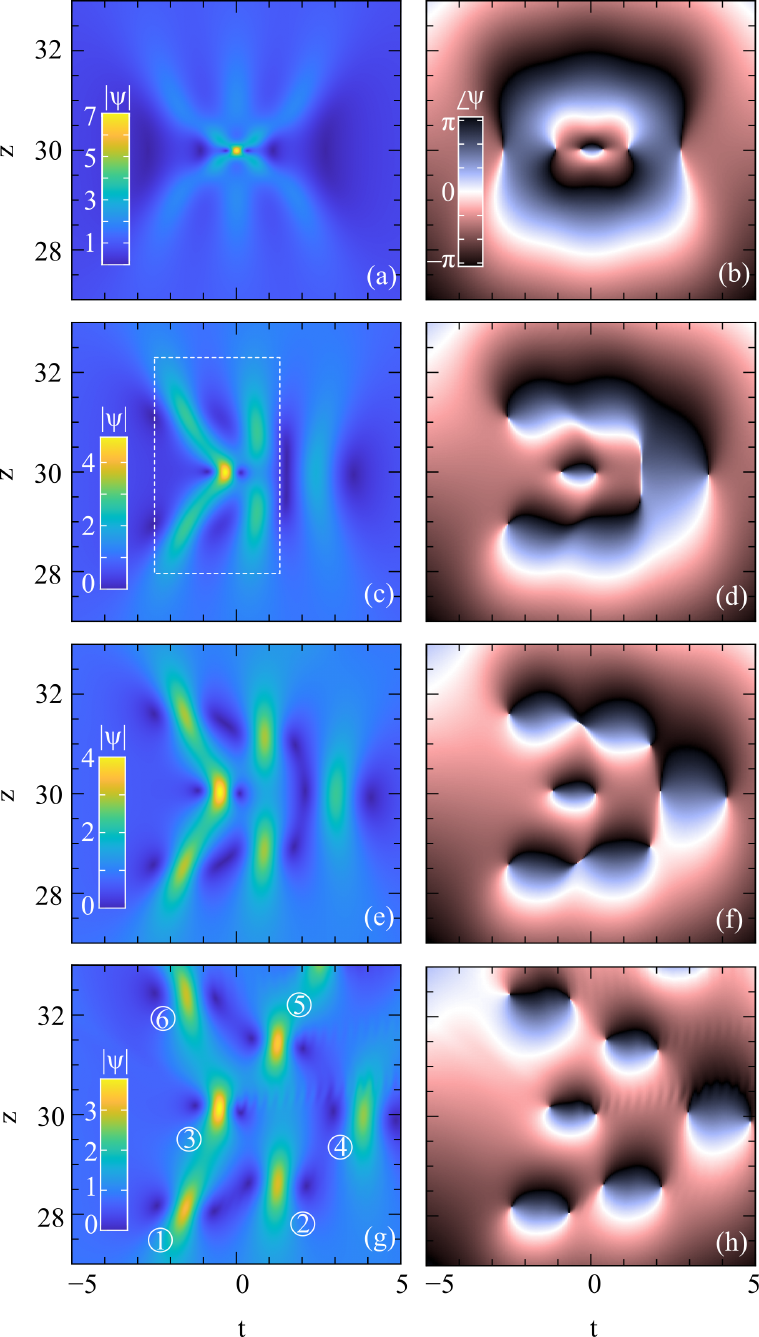}
\caption{(a) Amplitude and (b) phase of a third-order rogue wave with $\beta_3=0$. (c) and (d) show the onset of a separated second-order rogue wave along with a first-order one when $\beta_3=0.04$. Panels (e) and (f) show how the second-order rogue wave is transformed into first-order rogue waves when $\beta_3=0.08$. Finally, (g) and (h) show a cluster of well-separated first-order rogue waves at $\beta_3=0.15$. }
\label{third-fission}
\end{figure}

The same concept can be extended to $N$-th order rogue waves. We apply the perturbation $\beta_3$ to a third-order rogue wave, which, from the exact analytic solution, is known to consist of six fundamental rogue waves. We take the same numerical approach as in the previous sections. To observe the impact of the TOD, we varied $\beta_3$ from $0$ to $0.15$ and show selected cases in Fig.~\ref{third-fission}. Without any perturbation ($\beta_3=0$) the temporal and phase profile of a third-order rogue wave is presented in Fig.~\ref{third-fission}(a) and \ref{third-fission}(b). The amplitude is $7$ and the $\pi$ phase shift across the maximum height is visible at the centre of the phase profile. 

If we perturb the third-order rogue wave weakly ($\beta_3=0.04$, in Fig.~\ref{third-fission}(c)), the rogue wave breaks apart in two with a transient appearance of a second-order rogue wave accompanied with an isolated first-order rogue wave. The former (highlighted in figure by a white rectangle) reaches the maximum amplitude of $\approx 5$. The presence of $\beta_3$ makes its appearance highly distorted. The associated phase change is shown in Fig.~\ref{third-fission}(d). As the magnitude of the perturbation increases, ($\beta_3=0.08$) the short-lived second-order rogue wave also breaks apart into five premature first-order rogue waves with varying amplitudes, as shown in Fig.~\ref{third-fission}(e). Their phase profiles appear more and more independently with a phase shift of $\pi$ along the maximum amplitude, as the TOD $\beta_3$ increases (see Fig.~\ref{third-fission}(f)). 

This clearly demonstrate that the disintegration of higher-order rogue waves due to a weak TOD occurs via progressive separations following a hierarchical pattern. As such a second-order rogue wave produces a doublet and a first-order rogue wave. Similarly, a third-order rogue wave breaks into a short-lived second-order rogue wave and a first-order one. This pattern of disintegration is followed by any higher-order rogue waves under the effect of a small $\beta_3$.

Finally, when relatively strong perturbation ($\beta_3=0.15$) is applied to  the third-order rogue wave, the separation distance among the break-away components increases and the whole structure turns into a cluster of six fundamental rogue waves (marked in numbers and shown in the temporal profile in Fig.~\ref{third-fission}(g)). The break-away first-order rogue waves are completely independent from each other can be seen in the phase profile in \ref{third-fission}(h). Similar disintegration schemes applies to all $N$-th order rogue waves.

\begin{figure}[h]
\centering
\includegraphics{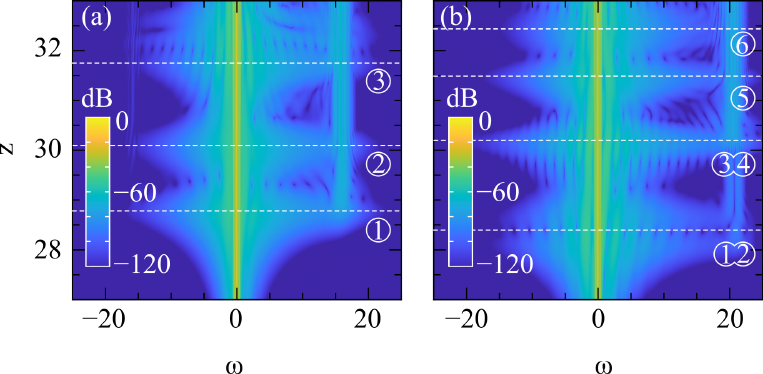}
\caption{(a) and (b) are the spectral profile of a second and third-order rogue wave with $\beta_3=0.20$ and $\beta_3=0.15$ respectively. Fig.~\ref{todsecond}(g) and Fig.~\ref{third-fission}(g) are their time-domain profiles. Each break-away rogue wave is accompanied with dispersive wave emission.}
\label{DS}
\end{figure}

One significant consequence of perturbing the higher-order rogue wave with TOD effect is dispersive wave (DW) emission. When the TOD effect is strong, each disintegrated rogue wave generates a group of linear waves which is phase matched with the rogue wave itself. In the spectral domain, the frequency components corresponding to these linear waves or DW at the maximally compressed points, 
becomes clear in Figs.~\ref{DS}(a) and \ref{DS}(b) respectively (their corresponding time domain images are Figs.~\ref{todsecond}(g) and \ref{third-fission}(g)). 

For second-order rogue waves (see Fig.~\ref{DS}(a)), each generated fundamental rogue wave produces the broadest spectrum at the maximum compression point, as indicated by the dashed white line. The corresponding rogue waves are marked with number within circles. Each of the rogue wave generates DW around $\omega \approx 20$ which is clearly visible that appear as a shoulder of component frequencies.

A similar effect appears in Fig.~\ref{DS}(b), which shows the spectral profile of the rogue wave cluster formed in Fig.~\ref{third-fission}(g). The spectra from rogue wave $1$ and $2$ are superimposed on each-other and produce a periodic spectral pattern along the bottom white-dashed line. The generated phase matched DW from both of these rogue waves creates a shoulder at $\omega \approx 20$. The second lowest dashed-white line in Fig.~\ref{DS}(b) shows the overlapping frequency spectrum of rogue waves $3$ and $4$, again forming a strong interference pattern and a shoulder at the edge of the spectrum showing DW emission. The spectrum of the rogue waves $5$ and $6$, on the other hand, remain sufficiently separated and their spectra do not mutually interfere. Their relative positions marks the end of the cluster and generate DW separately (marked by the two top white lines). The phase matched frequency, $\omega_{\textrm{DW}}$, can be approximated to the first order to be $\omega_{\textrm{DW}}=3/\beta_3$. The numerically observed radiations match closely with the earlier reports made in \cite{baronio2020resonant,chowdury2021rogue}.

\section{Effect of self-steepening}
\label{four}

By including the SS effect only with the NLSE, the Eq.~\ref{gnlse} becomes
\begin{equation}
\label{gnlse-ss}
i \frac{\partial \psi}{\partial z}-\frac{\beta_2}{2} \frac{\partial^2 \psi}{\partial t^2}+\gamma \,\psi\lvert\psi\rvert^2+is\frac{\partial}{\partial t}(\psi\lvert\psi\rvert^2)=0
\end{equation}
where $s$ is the coefficient of SS effect. Analytic first-order rogue wave solution of Eq.~\ref{gnlse-ss} is presented in \cite{chen2016chirped} with a complicated form. 
\begin{figure}[htbp]
\centering
\includegraphics{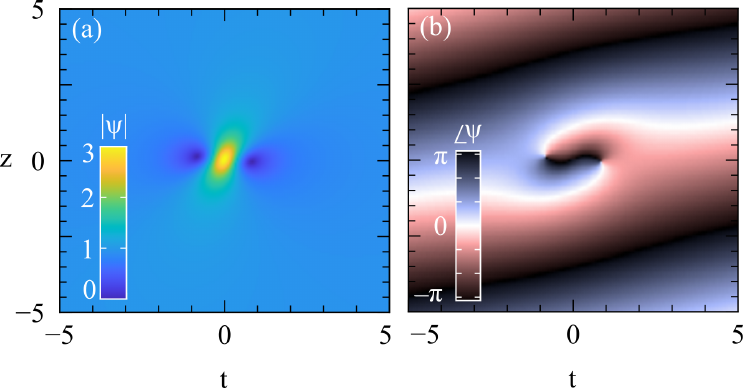} 
\caption{(a) Amplitude and (b) phase profiles of the rogue wave from Eq.~\ref{gnlse-ss} with $s=0.2$. It has the maximum amplitude of $3$ with a distorted phase shift of $\pi$ across the peak. Note that this solution formed at $t=z=0$ because of the closed form nature of the analytic solution.}
\label{steep}
\end{figure}

We re-formulate the solution into a simpler form which is given as:
\begin{equation}
\label{eq9}
\begin{aligned}
\psi_{s}(z,t)&= \left(1-\frac{G+ i H z+8 is \tau}{D_s}\right)e^{ i \left[ z \left(1+\frac{1}{2}s ^2\right)-t s+\Phi \right]}\textrm{,}
\end{aligned}
\end{equation}
\begin{figure}[h]
\centering
\includegraphics{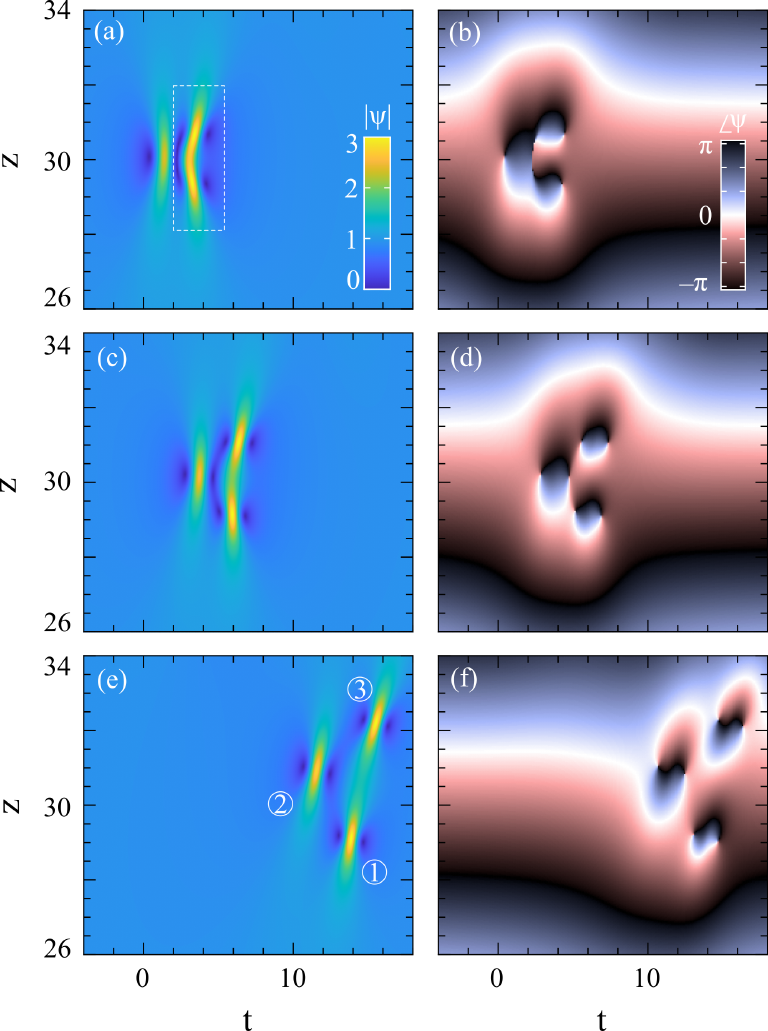}
\caption{A second-order rogue wave under the influence of SS effect. (a) temporal (b) phase profile with $s=0.01$ shows a doublet and a separated first-order rogue wave. Similarly, (c) and (d) are with $s=0.08$ showing the doublet separated in two first-order rogue wave amplitude reaching to $3$. (e) and (f) showing the disintegrated well separated first-order rogue wave with a higher magnitude of $s=0.2$. }
\label{sssecond}
\end{figure}
where $\tau=t-z s$, $\kappa=1+s^2$ and 
\begin{equation*}
\begin{aligned}
 D_s&=D+4 i s  (2 \tau -t)+4 s  \tau  (s  \tau -2 z)\textrm{,}\\
 D&=1+4t^2+4z^2\textrm{,}\\
 \Phi &= 2 \tan ^{-1}\left[\frac{4 s  (z s -\tau )}{1+4 \kappa  \left(z^2+\tau ^2\right)}\right]\textrm{.}\\
 \end{aligned}
\end{equation*}
Here, $\beta_2=-1$ and $\gamma=1$, while $s$ can be an arbitrary value. With $s\to0$ it directly reduces to fundamental rogue wave solution \cite{Kibler:10:Nature}. This solution profile can now be translated to any point on the $z$-$t$ plane following the relations $t=t_1-t_{s}$ and $z= z_1-z_{s}$. The SS effect induces a drift velocity in the development of the rogue wave along with a distorted time varying phase profile presented in Fig.~\ref{steep} (a) and \ref{steep}(b) respectively. This make the rogue wave to appear tilted at its emerging point. Depending on the magnitude of $s$, this tilted position varies. We shall see that the SS effect induces fission in higher-order rogue waves and the disintegrated first-order components display the same temporal and phase characteristics as in the Fig.~\ref{steep}.
 
Taking the exact analytic NLSE second-order rogue wave solution as initial condition, we numerically solved Eq.~\ref{gnlse-ss} for a range of $s$ values. We find that in the presence of SS, the second-order rogue wave experiences fission, breaking apart the structure shown in Fig.~\ref{todsecond}(a) into three fundamental rogue waves. Most importantly we observe that along with a distorted time varying phase profile, the SS effect also relocates the disintegrated rogue waves in a transversely shifted location. This indicates that in numerical simulation the SS effect spontaneously activates the translational parameter $t_s$ that arise in the analytic solution Eq.~\ref{eq9}.

When the SS coefficient $s=0$, the growth of MI for each of these fundamental rogue waves takes place at the same space and time in a synchronized way resulting in the appearance of a bound-state formation of a second-order rogue wave as shown in Fig.~\ref{todsecond}(a). However, the presence of the SS effects breaks this degeneracy allowing a space-time varying MI developemnt. Because of the SS effect while the rogue waves are developing, each of them experiences translation relatively different from each-other in the positive $t$ direction. The intrinsic translation among the disintegrated rogue wave components inhibit the formation of a second-order rogue wave, instead the three fundamental rogue waves appear in three distinct positions and times. 

SS-drive and TOD-driven fission share several similarities, with some notable exception. For instance, the breaking direction of the second-order rogue waves for TOD and SS effects are opposite to each other when the corresponding coefficients are both positive. As seen from Fig.~\ref{todsecond}(c), for TOD case the doublet has appeared in the left, where as with the SS effect with $s=0.03$, it appeared to the right at a $t$ translated location, as shown in Fig.~\ref{sssecond}(a). Using a relatively strong SS effect  ($s=0.06$), during the rogue waves development, the SS effect continuously shifts the component rogue waves. As a result, shown in Fig.~\ref{sssecond}(a), the conjoined doublet in the right is now broken apart in Fig.~\ref{sssecond}(c). The separation is also evident in the phase profile \ref{sssecond}(d). Such translation mechanism is absent in the case of TOD, where fissioned components emerge at around $\psi(z=30,t=0)$.

With even stronger value of $s=0.2$, the disintegrated rogue waves are far apart from each-other, and appear at three distinct positions and times, as shown in the temporal evolution (Fig.~\ref{sssecond}(e)) with a distorted phase profile  (Fig.~\ref{sssecond} (f)). The respective position of the fundamental rogue wave components $1$, $2$ and $3$ are opposite compared to the case of TOD induced fission. With further increased value of $s$, the transverse translation distance for the appearance of the rogue wave also grows. Note that the phase profile of each of the separated rogue wave closely match with that of the first-order rogue wave shown in Fig.~\ref{steep}(b).

\begin{figure}[h]
\centering
\includegraphics{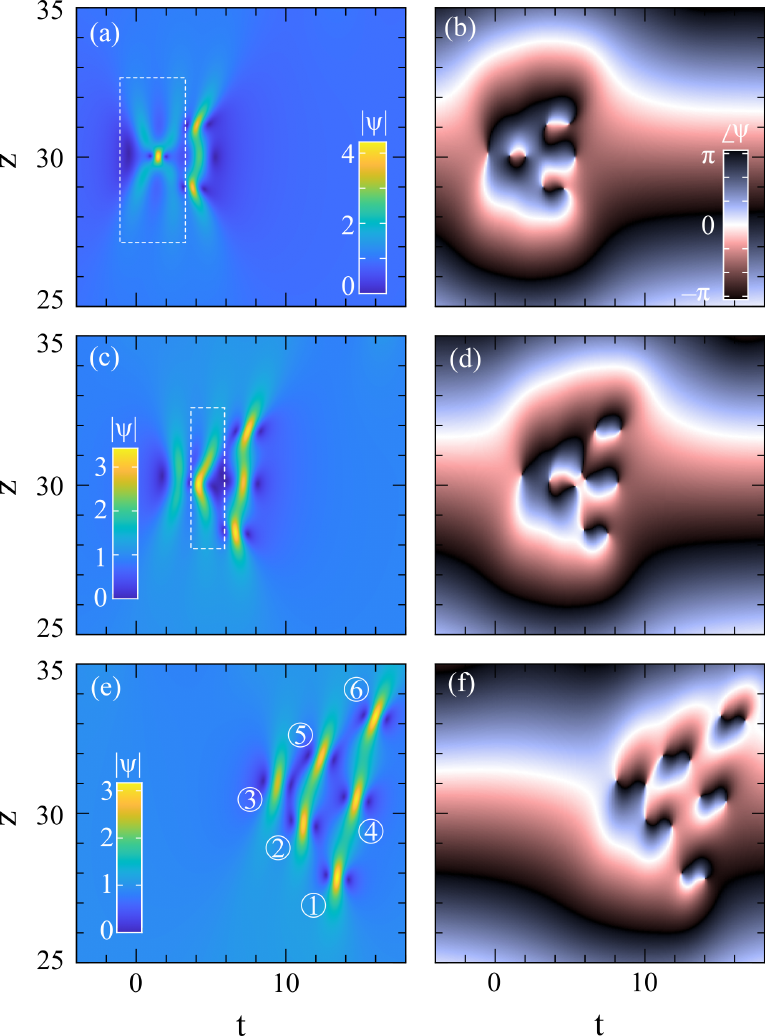}
\caption{(a) temporal (b) phase profile of a third-order rogue wave with $s=0.01$ showing the onset of a second-order rogue wave within the white rectangle appeared with three premature first-order rogue wave. (c) and (d) shows the second-order rogue wave now breaks apart with a doublet within the white rectangle along with fundamental rogue wave with $s=0.06$. (e) and (f) demonstrating the complete break-off of the third-order rogue wave reduced to well-separated six first-order rogue wave with comparatively strong value of $s=0.2$.}
\label{ssteepthird}
\end{figure}

The effect of SS on a third-order rogue wave is similar to that on the second-order one. To observe the process of fission we plot few examples. Interestingly, with a small value of $s=0.04$ the onset of a second-order rogue wave within the white rectangle is clear along with a group of three underdeveloped first-order rogue waves in the left (Fig.~\ref{ssteepthird}(a) and \ref{ssteepthird}(b)). The emerging point is slightly translated towards positive $t$. In  Figs.~\ref{ssteepthird}(c) and \ref{ssteepthird}(d) ($s=0.04$), this second order rogue wave is disintegrated into a doublet (encapsulated within a white rectangle) associated with a first-order rogue wave in the left and three fully developed rogue waves to the right. 

The breaking of the second-order rogue wave mimics exactly the same process as demonstrated in Fig.~\ref{sssecond}. However, with a higher-value of $s=0.2$, the transient appearance of the second-order rogue wave shown in Fig.~\ref{ssteepthird}(a) completely breaks apart into three fundamental rogue waves $2$, $3$ and $5$ in Fig.~\ref{ssteepthird}(e). In total the six rogue waves in Fig.~\ref{ssteepthird}(e) are well-separated and arranged in an asymmetric way, yet located at new transversely shifted locations. Note that the phase profile of each rogue waves in Fig.~\ref{ssteepthird}(e) is also distorted similarly to that of in Fig.~\ref{steep}(b).

\begin{figure}[h]
\centering
\includegraphics{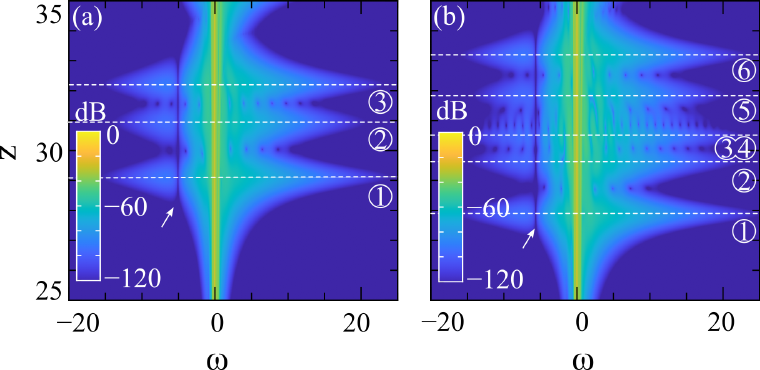}
    \caption{(a) and (b) are the spectral profile of Fig.~\ref{sssecond}(e) and Fig.~\ref{ssteepthird}(e) with self-steepening effect $s=0.2$ for both. Corresponding Spectral width of each of the rogue wave is shown in numbers.}
\label{ssfreq}
\end{figure}

In the spectral domain, each of the fissioned rogue wave develops optical chirp, resulting from the time varying phase development during the evolution under the SS effect. Optical chirp is a process where spectral density increases or decreases with time \cite{Agrawal:12:Book}. The origin of this instantaneous change comes from the SS term which in the Fourier domain becomes $\frac{\partial}{\partial t}(\psi\lvert\psi\rvert^2)= -i \omega (\psi\lvert\psi\rvert^2)$ by replacing the derivative ${\partial}/{\partial t}= -i \omega$. This results in an instantaneous asymmetrical spectral broadening in the rogue waves spectra. In Fig.~\ref{ssfreq}(a) three distinct asymmetrical spectra $1$, $2$ and $3$ correspond to the three numbered rogue waves formed in Fig.~\ref{sssecond}(e). Note that because of the time-varying distorted phase development, there is a spectral discontinuity in the spectral profile indicated by the white arrow in Fig.~\ref{ssfreq}(a).

Similarly, the spectra in Fig.~\ref{ssfreq}(b) correspond to the six disintegrated rogue waves presented in Fig.~\ref{ssteepthird}(e). In this figure, the distinct spectral profile appeared for rogue wave $1$, and $6$ only. The spectral components arise from the rogue waves $2$ and $5$ are partly interacting with $3$ and $4$ giving rise to an interfered spectral profile for both of the rogue wave $2$ and $5$. However, the spectral components arising from rogue waves $3$ and $4$ are mutually interacting and generating an overlapping spectral profile along the white dashed line shown in Fig.~\ref{ssfreq}(b). Similar to that of second-order case, a spectral discontinuity also arise here marked by a white arrow. Note that while there is a DW emission during the splitting in TOD case, is completely absent in the case of SS effect.

\section{Effect of RIFS}
\label{five}
 
The study of Raman effect on rogue waves has, so far, been done mostly on the first-order rogue wave \cite{ankiewicz2018rogue,ankiewicz2013rogue}. Similar studies on higher-order rogue waves has never been reported before. We use the equation:
\begin{equation}
\label{raman-eq}
i \frac{\partial \psi}{\partial z}-\frac{\beta_2}{2} \frac{\partial^2 \psi}{\partial t^2}+\gamma \,\psi\lvert\psi\rvert^2-\tau_R \psi \frac{\partial|\psi|^2}{\partial t}=0,
\end{equation}
\begin{figure}[ht]
\begin{center}
\includegraphics{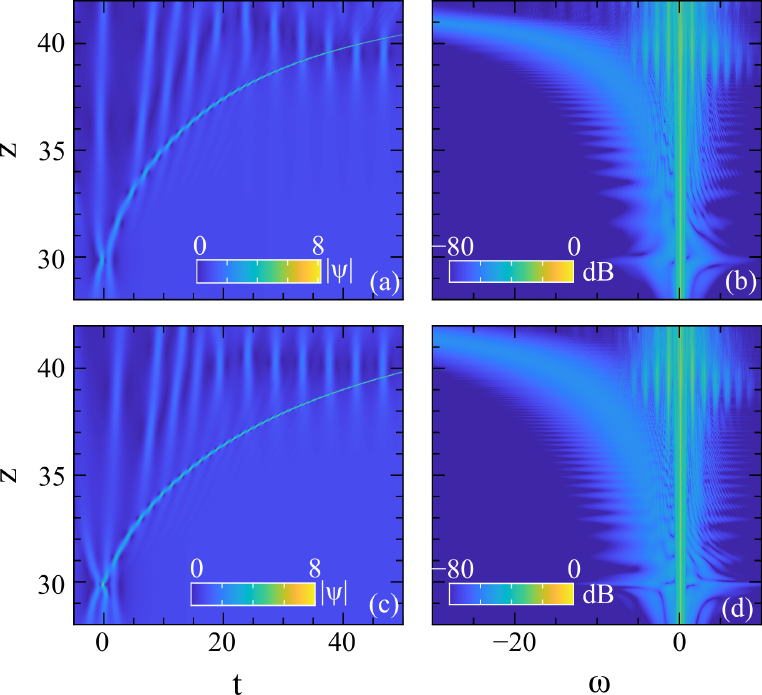}
\caption{ Impact of Raman effect on a second-order rogue wave. (a) the temporal evolution showing the ejection of an decelerated pulsating soliton. (b) is the spectral domain showing the Raman gain is generating the red-shifted spectral components with the expense of the blue spectral frequencies. The magnitude of the Raman coefficient is $\tau_R = 0.008$. (c) is the temporal profile a third-order rogue wave with Raman effect with $\tau_R=0.005$ showing the ejection of an decelerated soliton (d) is the spectral domain illustrating the red-shifted frequency generation.}
 \label{ramanfreq}
\end{center}
\end{figure}
to study the RIFS effect on higher-order rogue waves. Unlike the TOD and SS effect, the Raman term is a non-Hamiltonian, meaning that it does not preserve the energy of the system \cite{menyuk1993soliton}, and hence do not have an analytic solution. As a result when the rogue wave is evolving in $z$ it dissipates energy, thus altering its amplitude and width while shifting its central frequency.
 
Under weak RIFS effect, the disintegration of higher-order rogue wave follows the similar steps as for TOD and SS effects. For instance, we applied $\tau_R=0.008$ on a second-order rogue wave and observed its disintegration into a doublet (left) and a first-order rogue wave  (right) at the position of $\psi(z=30,t=0)$ (Fig.~\ref{ramanfreq}(a)). The separated first-order rogue wave immediately assumes the flight-trajectory of a soliton and gradually slows down. Along the path, while it decelerates in the positive $t$ direction, it emits red-shifted frequencies, as presented in Fig.~\ref{ramanfreq}(b). In other words, at this stage, because of the non-Hamiltonian nature of the solution, the rogue wave is no longer robust, but instead it loses energy by generating red spectral components. As the energy dissipation continues, the rogue wave slows down and decelerates assuming a bend trajectory, which is shown in Fig.~\ref{ramanfreq}(a).  
\begin{figure}[ht]
\begin{center}
\includegraphics{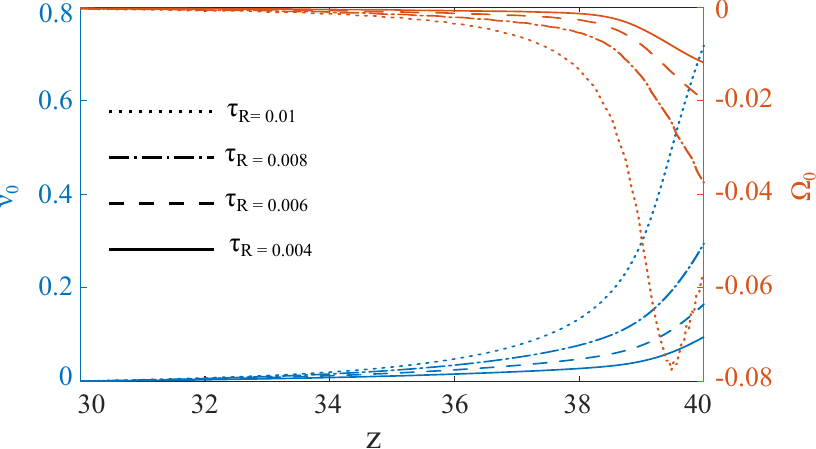}
\caption{Evolution of the center of mass of a second-order rogue wave. The blue curves are the temporal and red curves are the spectral evolution. As there is no development before $z=30$, the advanced evolution is shown from $z=30$ to $z=40$.}
 \label{cmass}
\end{center}
\end{figure}
Similar behaviours are observed for a third-order rogue wave as shown in Fig.~\ref{ramanfreq} (c) and Fig.~\ref{ramanfreq} (d). Note that with an applied RIFS effect with $\tau_R=0.005$ on a third-order rogue wave, a transient appearance of a second-order rogue wave also observed, similarly to the case of TOD and SS effects (Fig.~\ref{third-fission}(c) and Fig.~\ref{ssteepthird}(a) respectively).

The energy dissipation and the subsequent change in amplitude and frequency shift can be described by the progression of the center of mass of the rogue wave while it is under the influence of RIFS effect. We define the center of mass of the evolving rogue wave in temporal $\nu_0$ and spectral domain $\Omega_0$ as
\begin{equation}
\nu_0 = \frac{\int_{-\infty}^{\infty}\,t\,|\psi(z,t)|^2 dt}{\int_{-\infty}^{\infty}|\psi(z,t)|^2 dt}, \;\;\Omega_0 = \frac{\int_{-\infty}^{\infty}\,\omega\,|\psi(z,\omega)|^2 d\omega}{\int_{-\infty}^{\infty}|\psi(z,\omega)|^2 d\omega}
\end{equation}
\begin{figure}[ht]
\begin{center}
\includegraphics{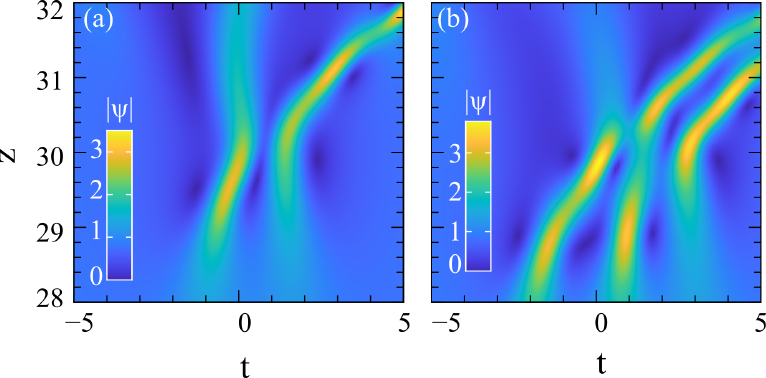}
\caption{(a) the fission of a second-order and (b) a third-order rogue wave when under the influence of a strong Raman effect with $\tau_R=0.20$ and $0.25$ respectively.}
 \label{rfission}
\end{center}
\end{figure}

Fig.~\ref{cmass} shows how $\nu_0$ and the $\Omega_0$ of a pulsating rogue wave is evolving for four different values of the coefficient $\tau_R=0.004$, $0.006$, $0.008$ and $0.01$. From the trajectory of $\nu_0$, it is clear that for high value of $\tau_R$, the rogue wave advances more slowly indicating higher energy leakage. This induces a more skewed bow-shape trail. Due to the red-shifted frequency emission, the center of mass of the frequency profile shifted towards negative $\Omega_0$. With increased evolution distance, as the rogue wave continues to lose energy, the pulsating rogue wave becomes more compressed with a wider bandwidth spectral profile. As a result, more shifting of the centre of mass of the frequency profile towards negative $\Omega_0$ direction occurs.

Under the RIFS effect, the full disintegration of higher-order rogue waves requires stronger $\tau_R$ values. As shown in Fig.~\ref{rfission}(a), a second-order rogue wave disintegrated into three fundamental rogue waves with $\tau_R=0.20$. With such strong RIFS value, the separated rogue wave distorted significantly because of strong decelerating effect. Similar observation also noted in the case of a third-order rogue wave shown in Fig.~\ref{rfission}(b), where it disintegrated into six fundamental rogue waves with $\tau_R=0.25$.

\begin{figure}[ht]
\begin{center}
\includegraphics{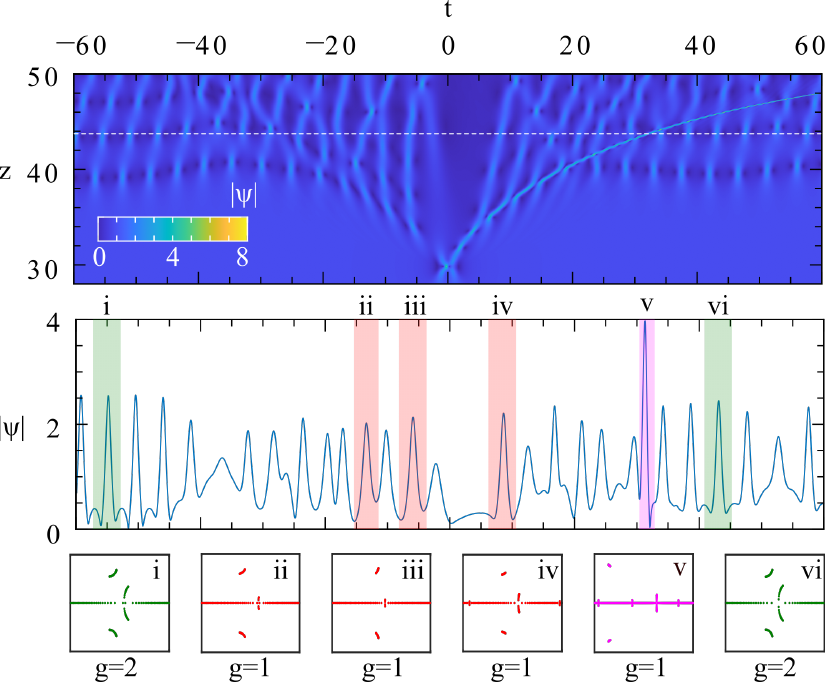}
\caption{Top panel is the temporal profile of the evolution of a rogue wave under the influence of RIFS effect. The rogue wave developed at $z=30$, and a skewed background wave dynamics is visible with a combination of different types of wave entities. In the mid-panel, the separated wave envelop is an instance that extracted at $z=43.8$ showing with dashed-white line. Bottom panel shows the IST spectrum of various types of wave profile within a box. Note that the top and mid panel shares the same $t$ axis.} 
 \label{fig_IST}
\end{center}
\end{figure}

Higher-order rogue waves also influence the neighbouring background when it is under the effect of RIFS. To investigate this, we simulate a second-order rogue wave with $\tau_R=0.004$ for an extended period of time shown in the top panel in Fig.~\ref{fig_IST}. Clearly, the continuous wave background is now highly distorted with a variety of other waves appearing in the neighbourhood of the evolving rogue wave. To classify the types of waves formed, we take an envelop at $z=43.8$ shown with a white dashed line in the top panel. Its corresponding amplitude plot is shown in mid-panel. We select six representative amplitude profiles from the envelop as sample and employ the IST-spectral analysis \cite{randoux2016inverse}. 

The IST reveals a combination of spectral bands in these chosen localised structures shown within six-boxes in the bottom panel. In the mid-panel, the structures within the green-shaded areas $\text{i}$ and $\text{vi}$ have three spectral band, making them genus-2 solution which allows to classify them to be either breathers or rogue waves. Similarly, the localized structures $\text{ii}$, $\text{iii}$ and $\text{iv}$ within the pink-shaded areas present two-spectral band, and therefore can be classified as solitons (which belongs to genus-1-type solutions).  Finally, the localised structure in the magenta-shaded area, is the soliton that is created from the decelerating rogue wave. The corresponding eigenvalues for this structure (box-$\text{v}$) are not centered around the real line. This occurs because the shape of the corresponding soliton is highly asymmetric. These observation indicates that higher-order rogue waves under the influence of RIFS effect trigger the formation of a series of other waves around it, such as breathers and solitons. 

\section{Combined effect}
\begin{figure}[ht]
\begin{center}
\includegraphics{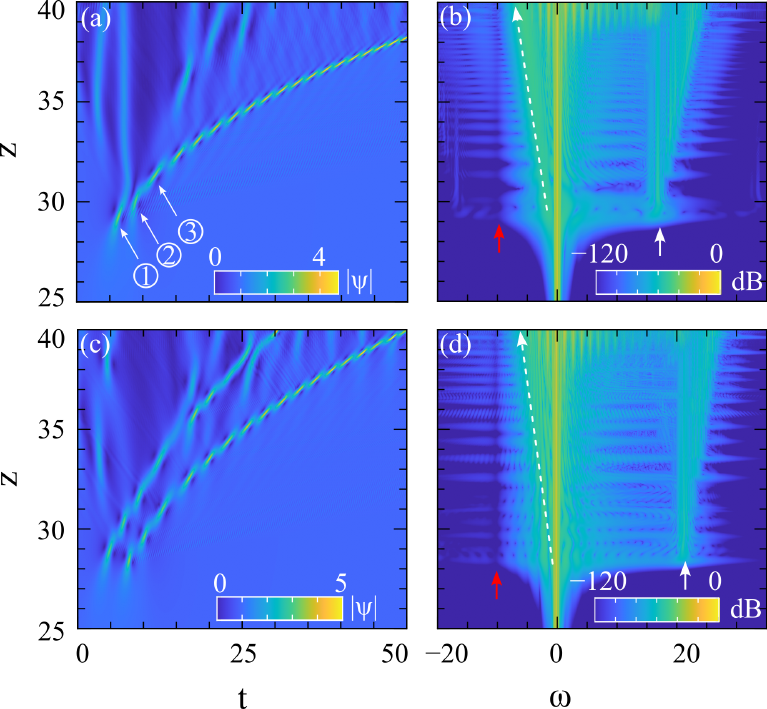}
\caption{Combined effects of TOD, SS and RIFS effects on a second and third-order rogue wave. (a) temporal (b) spectral evolution of a second-order rogue wave where the leading third disintegrated rogue wave component is accelerating with the simultaneous generation of blue and red-shifted frequency components. The coefficient values are $\epsilon_3= 0.2$, SS $ s=0.10$ and $\tau_R=0.008$. Similarly (c) temporal and (d) spectral domain of a disintegrated third-order rogue wave with the coefficient values as $ \epsilon_3= 0.15$, $s=0.10$, and $\tau_R=0.005$. It created two trails of decelerating rogue waves generating blue and red shifted frequency components.}
 \label{final}
\end{center}
\end{figure}

We now address the case where all the three perturbations, TOD, SS and RIFS, are present. We find that the process of rogue wave fission persists. Fig.~(\ref{final})(a) shows temporal evolution of a second-order rogue wave fission. We observe that all the disintegrated components assume the same trajectory towards the positive $t$ direction. 

As noted earlier, due to TOD effect, the orientation of the breakaway first-order rogue waves from a second-order one is opposite to that of SS effect shown in Fig.~\ref{todsecond}(g) and Fig.~\ref{sssecond}(e). This opposite ejection acts as a balance, aligning them when they are in a combined effects. Though the RIFS effect induced fission in the main structure is similar to that of TOD, however, its leading mechanism is to slowdown the appeared rogue waves by leaking energy from them.

\label{six}

Note that the RIFS effect is inversely proportional to the pulse duration $t_0$ and a rogue wave's pulse duration must be short enough to come under the active influence of RIFS effect. In Fig.~\ref{final}(a), component $1$ and $2$ remain stationary because their durations are not within the range to be affected by the RIFS effect. However rogue wave $3$ is at the leading position and achieve comparatively shorter pulse duration that enable it to come under the active influence of RIFS effect resulting in a bent trajectory.

\begin{figure}[ht]
\begin{center}
\includegraphics{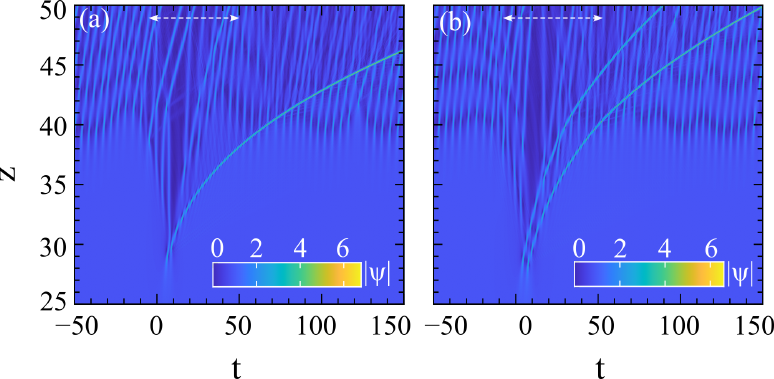}
\caption{(a) shown from the extended evolution of Fig.~\ref{final}(a), the decelerating rogue wave from a disintegrated second-order rogue wave transformed into a soliton. (c) is the prolonged evolution from Fig.~\ref{final}(c) for a third-order rogue wave. It generates two solitons. However, for both case, low amplitude solitons are also ejected from around the central region marked by white arrows. }
 \label{final2}
\end{center}
\end{figure}

The dynamics become clearer in the spectral domain (see Fig.~\ref{final}(b)). The fixed components $1$ and $2$ do not lose energy, as a result they do not generate red-shifted frequency components. Nevertheless, due to the presence of TOD effect they generates DW indicated by the white arrow. On the contrary, rogue wave $3$ with shorter pulse duration is under the active influence of RIFS effect and continues to lose more energy as it evolves further. Along the evolution, the repeated compression stages of rogue wave $3$ creates a cascaded emission of DW. As the rogue pulse becomes highly compressed, it achieve enough spectral bandwidth to pump and create the red-shifted frequency components as well. A steady red-shifted skewed development of frequencies is clearly visible within the frequency range $\omega = 0$ to $-10$ indicated by the long white arrow in Fig.~\ref{final}(b). Note that a spectral discontinuity is observed due to the time varying phase development related to spectral chirp due to SS effect. It is also responsible for the achieved spectral asymmetry.

We have also investigated the combined effects on a third-order rogue wave. TOD effects tend to break the third-order rogue wave apart and arrange its components in a cluster as shown in Fig.~\ref{third-fission}(g). On the other hand, the SS effect tends to arrange them in a triangular fashion, as shown in Fig.~\ref{ssteepthird}(e). When both of these effect act on a third-order rogue wave simultaneously, together with RIFS effect, they counter balance the directional arrangements of the disintegrated rogue wave and force its components to align in a parallel fashion as shown in Fig.~\ref{final}(c). This is similar to the case of a second-order rogue wave described above. The Raman effect eventually slows-down this parallel arrangement in the forward evolution direction, generating both blue and red-shifted frequency components (indicated by the long white arrow). Notice that SS induced spectral discontinuity indicated by red arrow and  DW emission indicated by white arrow are also present.

As the system keep evolving, the rogue waves eventually transform into a group of fundamental solitons. This is shown in Fig.~\ref{final2}(a) (extended evolution of Fig.~\ref{final}(a)) for a second-order rogue wave. After fission, it triggers a collection of low-amplitude solitons created around the central region (indicated by the white horizontal arrow). The rogue wave itself, now transformed into a fundamental soliton, proceed with a bow-trajectory. Similarly, in Fig.~\ref{final2}(b), a third-order rogue wave also eject a small number of solitons in the central region. This time however,  it transforms into two fundamental solitons. In both cases, a breathers-type formation emerges near the edge of the evolution field due to MI on an unstable wave background.

\begin{figure}[ht]
\begin{center}
\includegraphics{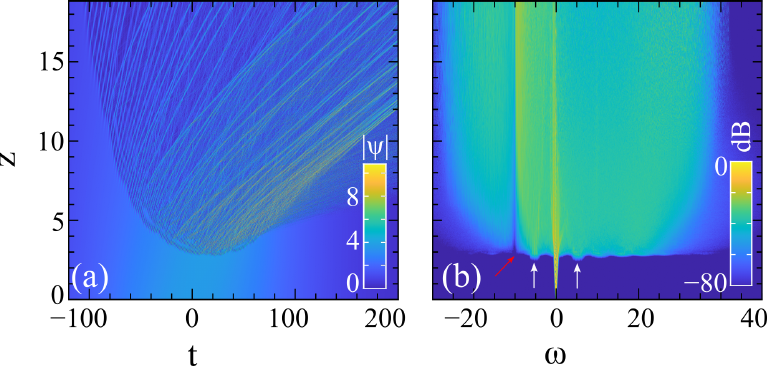}
\caption{(a) temporal and (b) spectral domain presentation of the MI induced disintegration of $N=300$ soliton under the influence of TOD, SS and RIFS effects with $s=0.1$, $\epsilon_3=0.03$ and $\tau_R=0.001$ respectively. Onset of MI is shown with white arrows while the spectral discontinuity arises due to SS effects shown with a red arrow.}
 \label{final3}
\end{center}
\end{figure}

It is possible to show that, this type of rogue waves formation, fission and subsequently their soliton transformation has real world implication. For example, in CW-SCG, it is observed that, initially higher-order soliton goes through MI and the end product is hundreds of solitons \cite{russell2014hollow,travers2011ultrafast}. To show this we simulated $N=300$ soliton using Eq.~\ref{gnlse} under the influence of TOD, SS and RIFS effects. The evolution dynamics in temporal domain is presented in Fig.~\ref{final3} (a). It shows that initially, the noise driven MI leads to the formation of hundreds of rogue wave-type substructures that eventually transforming to many solitons. The onset of MI is clearly visible in the spectral domain in Fig.~\ref{final3}(b) with the appearance of side-lobes around $\omega=0$ as an evidence of MI development.

A large number of soliton together can be act as a continuous wave background. Presence of noise among the solitons can spontaneously trigger the MI which leads to the development of both fundamental as well as higher-order rogue waves on it. This type of formations as a result of noise seeded MI is reported in \cite{toenger2015emergent}. If the formation is a fundamental rogue wave, the concurrent influence of TOD, SS and RIFS effect transforms it directly to a small number of those solitons shown in Fig.~\ref{final3}(a). However if MI contributes to the development of higher-order rogue waves, first, the combined effects of TOD, SS and RIFS disintegrates them into a group of fundamental rogue waves, which eventually transforms into a bunch of solitons as shown in Fig.~\ref{final3}(a). 

Thus whether it is a fundamental rogue wave or a higher-order rogue wave that is formed in the initial stage of the MI, under the influence of TOD, SS and RIFS effects, the end product is always be a large collections of fundamental solitons.

\section{Conclusion}

Using extensive numerical approaches, we have showed that higher-order rogue wave undergo fission in a system weakly perturbed by the TOD, SS and RIFS effects. This is similar to the higher-order soliton fission. Under these effects, employing the second and third-order rogue wave solutions, we reveal their breaking mechanisms and how they reduce to their constituent parts. 

Importantly, we observed that, if the applied perturbation is weak, the higher-order rogue waves reveal their hierarchical pattern, such as a second-order rogue wave shows that it is made of one doublet and a first-order rogue wave. Similarly, a third-order rogue wave reveals that it is build on a second-order rogue wave together with three fundamental rogue waves. However, under strong perturbation they completely disintegrate to their constituent parts.

We observe that with the weak effect of RIFS, a higher-order rogue wave immediately ejects one or more decelerating fundamental rogue waves. Since the Raman effect is a dissipative term, the ejected rogue waves loses energy during its evolution, leading to a red shift in the frequency domain. With all the effects combined, we observe that after fission, the higher-order rogue wave trigger a collection of solitons. In the frequency domain this appear as asymmetrical new blue and red shifted spectral components. 

These new insights may provide a pathway to a new kinds of supercontinua, to produce frequency components beyond the traditional higher-order soliton based supercontinuum generation.  Moreover, these new observation may prove to be useful for interpreting various other nonlinear phenomena in optics, hydrodynamics and similar systems.

\section*{Acknowledgements}

AC and MB acknowledge Nanyang Technological University, NAP-SUG grant. WC acknowledges funding support from Ministry of Education--Singapore, AcRF Tier 1 RG135/20.

\bibliographystyle{unsrt}


\bibliography{Bibliography-2}

\end{document}